\documentclass[twocolumn,superscriptaddress,showpacs,floatfix,aps,prb]{revtex4}
\usepackage[dvips]{graphicx}
\usepackage{dcolumn}
\usepackage{amsmath}
\usepackage{amssymb}
\usepackage{amsbsy}
\usepackage{units}
\usepackage{color}

\newcommand{\ZMO}{{Mg$_{x}$Zn$_{1-x}$O}}
\newcommand{\ZNO}{{ZnO}}
\newcommand{\MGO}{{MgO}}
\newcommand{\phase}[1]{\ensuremath{\mathsf{#1}}}
\newcommand{\transition}[2]{\ensuremath{\mathsf{#1}\leftrightarrow\mathsf{#2}}}

\begin{document}
\title{Structural phase transitions and fundamental band gaps of
  Mg$_{x}$Zn$_{1-x}$O alloys from first principles}

\author{I. V. Maznichenko} 
\affiliation{Martin-Luther-Universit\"at Halle-Wittenberg, Fachbereich
  Physik, D-06099 Halle, Germany}

\author{A. Ernst}
\email{aernst@mpi-halle.de}
\affiliation{Max-Planck-Institut f\"ur Mikrostrukturphysik, Weinberg
  2, D-06120 Halle, Germany}

\author{M. Bouhassoune}
\affiliation{Max-Planck-Institut f\"ur Mikrostrukturphysik, Weinberg
  2, D-06120 Halle, Germany}
\affiliation{Department Physik, Universit\"at Paderborn, 33095 Paderborn, Germany}

\author{J. Henk}
\affiliation{Max-Planck-Institut f\"ur Mikrostrukturphysik, Weinberg
  2, D-06120 Halle, Germany}

\author{M. D\"ane}
\affiliation{Martin-Luther-Universit\"at Halle-Wittenberg, Fachbereich
  Physik, D-06099 Halle, Germany}
\affiliation{Materials Science and Technology Division, Oak Ridge
  National Laboratory, Oak Ridge, TN 37831, USA} 

\author{M. L\"uders}
\affiliation{Daresbury Laboratory, Daresbury, Warrington, WA4 4AD, UK}

\author{P. Bruno}
\affiliation{Max-Planck-Institut f\"ur Mikrostrukturphysik, Weinberg
  2, D-06120 Halle, Germany}
\affiliation{European Synchrotron Radiation Facility -- BP 220, F-38043
  Grenoble Cedex, France}

\author{W. Hergert}
\affiliation{Martin-Luther-Universit\"at Halle-Wittenberg, Fachbereich
  Physik, D-06099 Halle, Germany}

\author{I. Mertig}
\affiliation{Max-Planck-Institut f\"ur Mikrostrukturphysik, Weinberg
  2, D-06120 Halle, Germany}
\affiliation{Martin-Luther-Universit\"at Halle-Wittenberg, Fachbereich
  Physik, D-06099 Halle, Germany}

\author{Z. Szotek}
\affiliation{Daresbury Laboratory, Daresbury, Warrington, WA4 4AD, UK}

\author{W. M. Temmerman}
\affiliation{Daresbury Laboratory, Daresbury, Warrington, WA4 4AD, UK}

\date{\today}

\begin{abstract}
  The structural phase transitions and the fundamental band gaps of
  \ZMO\ alloys are investigated by detailed first-principles
  calculations in the entire range of Mg concentrations $x$, applying
  a multiple-scattering theoretical approach (Korringa-Kohn-Rostoker
  method). Disordered alloys are treated within the coherent potential
  approximation (CPA). The calculations for various crystal phases have
  given rise to a phase diagram in good agreement with experiments and 
other theoretical
  approaches.  The phase transition from the wurtzite to the rock-salt
  structure is predicted at the Mg concentration of $x=0.33$, which
is close to the experimental value of $0.33-0.40$.
  The size of the fundamental band gap, typically underestimated by the local
  density approximation, is considerably improved by the
  self-interaction correction. The increase of the gap upon alloying
  ZnO with Mg corroborates experimental trends.
  Our findings are relevant for applications in optical, electrical,
  and in particular in magnetoelectric devices.
\end{abstract}


\maketitle

\section{Introduction}
\label{sec:introduction}
In recent years, much effort has been devoted to research on ZnO,
inspired mostly by its attractive properties for optoelectronic
applications.\cite{Ozgur05} This interest arises from specific
properties, e.\,g.\ a large piezoelectric coefficient,
photoconductivity, and transparency in the visible and infrared
wavelength regimes. The range of applications of this semiconductor
can be considerably extended by alloying. Prominent dopants are Co and
in particular Mg, on which we focus in this work.  An increase of the
Mg concentration can transform the crystal lattice from the wurtzite
structure (\phase{WZ}) of ZnO to the rock-salt structure (\phase{RS})
of MgO\@.  Accompanied by this structural phase transition is a
substantial increase of the fundamental band gap.  The latter can be
tuned from \unit[3.35]{eV} to \unit[7.7]{eV}.\cite{Ozgur05,Chen06} In
view of a possible band-gap engineering, \ZMO\ alloys may also be considered
as suitable insulating spacers in magnetoelectronic devices, in
particular in magnetic tunnel junctions.

According to the equilibrium phase diagram,\cite{Segnit65,Raghavana91}
the solid solution of \ZMO\ is of eutectic type at normal conditions.
It is characterized by an extensive solubility of zincite in MgO (up
to \unit[33]{mol\%}) and by a restricted solubility of MgO in ZnO
(\unit[4]{mol\%}). The solubility depends strongly on experimental
conditions and can be considerably increased at high temperatures and
high pressures.\cite{Baranov05,Solozhenko06} Non-equilibrium growth
processes, like pulsed laser deposition
(PLD)\cite{Ohtomo98,Sharma99b,Choopun02,Narayan02,Kunisu04} and
molecular beam epitaxial methods\cite{Park01,Takagi03,Fujita04,Vashaei05}, allow to
grow high-quality \ZMO\ thin films for a large range of concentrations
$x$.  For \phase{RS}-\ZMO, produced by PLD, a maximum solubility has
been reported for $x = 0.5$.\cite{Choopun02,Bhattacharya03} When increasing
the Zn concentration, the \phase{RS} and the \phase{WZ} phases
separate, and \ZMO\ exhibits a \phase{WZ} structure for $x <
0.4$.\cite{Bendersky05} The solubility limit and the phase formation 
in \ZMO\ are strongly influenced by experimental conditions and by the
substrate on which the \ZMO\ film is grown.  In general, alloying of
ZnO and MgO proceeds by substituting Mg atoms by Zn atoms in the cubic
\phase{RS} structure and \textit{vice versa} in the hexagonal
\phase{WZ} structure.

The composition and the crystalline structure affect directly the
electronic properties of \ZMO\@. Numerous absorption and
photoluminescence spectroscopy experiments show that the fundamental
band gap depends differently on the Zn concentration in the \phase{RS}
and \phase{WZ} phases.\cite{Chen04,Schmidt-Grund05} The width of the
band gap increases most linearly with Mg concentration for both the
\phase{RS} and the \phase{WZ} phase, but the slope in cubic \ZMO\ is
about twice as large as in the hexagonal structure.\cite{Chen03}  This
experimental finding clearly indicates that \ZMO\ is a promising
candidate for band-gap engineering. For instance in a
magnetic tunnel junction, electrons that are transmitted from one
electrode to the other have to pass the nonconducting
spacer.\cite{Datta95} The transmission probability decays with spacer
thickness and with the width of the spacer's fundamental band
gap.\cite{Dederichs02,Zhang03c} As a consequence, the spin-dependent
conductance, i.\,e.\ the tunnel magnetoresistance, could be tuned by
varying the fundamental band gap. 
Further, the magnetoresistance
depends essentially on the properties of the ferromagnet-insulator
interface, as was shown for Fe/MgO/Fe tunnel
junctions.\cite{Zhang03d,Tusche05} Hence, detailed knowledge of its
geometric structure is necessary, and our theoretical investigation of
the bulk structural phases can be regarded as one step towards that
goal.

In addition to the extensive experimental work, there exist many 
detailed theoretical studies of both ZnO and MgO.
Among them are 
Refs.~\onlinecite{Jaffe93,Desgreniers98,Recio98,Hill00,Jaffe00,Limpijumnong01a,Limpijumnong01b,Baranov02,Sun05,Laskowski06,Uddin06}, using a variety of {\it ab initio} computational methods.
There are, however, not many studies of \ZMO\ alloys,
possibly due to the complex interplay of their electronic and geometric 
structures. Recently,
thermodynamical stability and ordering tendencies of the alloys have 
been carefully investigated using the cluster-expansion
method.\cite{Sanati03b} Based on the parameterization of total
energies for various alloy configurations, the latter allows to study
accurately structural properties, including short-range order (SRO)
effects.\cite{Zunger94,Laks92} It is found that the
\transition{RS}{WZ} transition occurs at $x \approx 0.33$, which is
consistent with experiment ($x = 0.33$, Ref.~\onlinecite{Ohtomo98};
similar results were obtained by a pseudopotential method using
supercells\cite{Kim01b}). The cluster expansion method was also
used by Seko \textit{et al.} for investigating phase transitions,
including vibrational effects through lattice dynamics
calculations.\cite{Seko05} The authors demonstrate that the transition
pressure decreases with increasing Mg content, which is explained as
follows. Below the solubility limit of MgO in ZnO, the \phase{RS}
phase is energetically preferred to the \phase{WZ} phase in MgO\@.
Above the solubility limit, the configurational entropy increases by
the transition from a mixed \phase{WZ}-ZnO/\phase{RS}-MgO to a
\phase{RS} single phase.

A brief critical review of previous theoretical
\textit{ab initio} investigations has shown that: (i) In ZnO, the energy
levels of the localized Zn-$3d$ electrons are found relatively high and thus
close to the valence bands, resulting in strong hybridization with O-$2p$
states. Since these hybridization effects have to be correctly taken into
account, the all-electron methods are preferred to the pseudopotential
methods in which localized electrons are neglected. (ii) Because the
Zn-$3d$ electrons are localized, they are not well described within
the local spin density approximation (LSDA) to density functional theory
(DFT). To treat these electronic states adequately, one has to go
beyond the LSDA, for example by applying the self-interaction
correction (SIC) to the LSDA\@. (iii) Previous studies often focused on
the \phase{WZ} and \phase{RS} structures of the ordered alloys
(i.\,e.\ $x = 0$ or $1$). A detailed investigation of the complete
transition path (with continuous variation of $x$) is still missing.
For the latter it is inevitable to treat disordered alloys, for
example within the coherent potential approximation.

In the present paper, we report on a systematic first-principles study
of structural and electronic properties of ordered and disordered
\ZMO\ alloys using an all-electron full-charge density
Korringa-Kohn-Rostoker (KKR) method.\cite{Korringa47,Kohn54} Alloying 
of MgO and ZnO is described within CPA,\cite{Soven67} as
formulated in multiple-scattering theory (KKR-CPA).\cite{Gyorffy72a}
One objective of this investigation is to describe accurately the
structural phase transition from the \phase{WZ} to the \phase{RS}
structure in \ZMO, by continuously increasing the Mg concentration.
The second objective addresses the formation of the fundamental band
gap in \ZMO\@. It is further shown how the hybridization between
Zn-$3d$ and O-$2p$ states affects the width of the band gap, by
comparing results obtained within the LSDA with those obtained by
applying the self-interaction correction.\cite{Perdew81,Lueders05} 
The validity of the present approach, especially the use of the CPA, is
discussed by comparing our results with those of previous
studies.\cite{Kim01b,Sanati03b,Seko05} In summary, our study addresses
important issues in \ZMO\ alloys which, with respect to magnetic
tunnel junctions, might also be relevant for magnetoelectronics.

The paper is organized as follows. The model of the transition path
from the \phase{WZ} to the \phase{RS} structure is sketched in
Section~\ref{sec:struct-phase-trans}. Details of the computational
approach are presented in Section~\ref{sec:comp}. Results are
discussed in Section~\ref{sec:struct-prop-order}. By comparing our
results for the ordered alloys with those of other theoretical work
and with experiment, the validity of our approach has been established.
Our main results for the disordered alloys are discussed in
Section~\ref{sec:struct-prop-disorder}. In
Section~\ref{sec:band-gap-engineering}, the formation and evolution of the
fundamental band gap as a function of Mg and Zn contents in the
system is analyzed. Concluding remarks close the paper
in Section~\ref{sec:conclusions-outlook}.

\section{Modeling the structural \transition{WZ}{RS} phase transition}
\label{sec:struct-phase-trans}
To investigate structural phase transitions in \ZMO, a transition path
suggested for the continuous structural \transition{WZ}{RS}
transformation in GaN was adopted.\cite{Limpijumnong01a} This path has
been already successfully applied in studies of structural
deformations in MgO\cite{Limpijumnong01b} and in
ZnO.\cite{Limpijumnong04} The structural transition is described as a
homogenous strain deformation from the \phase{WZ} to the \phase{RS}
phase by passing an intermediate hexagonal structure (\phase{HX},
refered to as $h$-MgO in Ref.~\onlinecite{Limpijumnong01a}).  
This hexagonal structure can as well occur in epitaxial systems due to the
reduction of the interlayer distance, which was recently observed for thin
ZnO films.~\cite{Tusche07,Meyerheim09} 
A similar scheme is the Bain's path for the transition
from the face-centered-cubic to the body-centered-cubic
structure.\cite{Marcus02}

\begin{figure}
  \centering
  \includegraphics[width = 0.95\columnwidth]{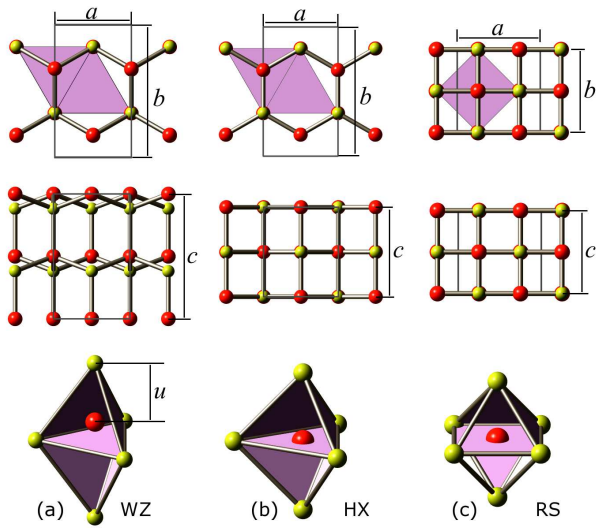}
  \caption{(Color) Structural phases in \ZMO\@. The
    wurtzite structure (a: \phase{WZ}, left column), the intermediate
    hexagonal structure (b: \phase{HX}, central column), and the
    rock-salt structure (c: \phase{RS}; right column) are shown in top
    view (top row) and in side view (central row). The hexahedral
    chemical units are depicted in the bottom row. The lattice
    parameters $a$, $b$, and $c$ for the common space group $Cmc2_{1}$
    are indicated. The $xy$-plane of unit cells for $P6_{3}mc$,
    $P6_{3}/mmc$, and $Fm\bar{3}m$ space groups for \phase{WZ},
    \phase{HX}, and \phase{RS} respectively are displayed in the top
    row in magenta. $u$ describes the internal displacement of cations
    sites (red, dark spheres) with respect to the central planes of
    the polyhedra. For $u = \nicefrac{1}{2}$, metal sites lie within
    the central planes which are spanned by the oxygen sites (yellow,
    bright spheres).}
  \label{fig:str}
\end{figure}

In the first step on the transition path (\transition{WZ}{HX}), the
internal parameter $u$ is linearly increased while simultaneously
decreasing the $\nicefrac{c}{a}$ ratio (Fig.~\ref{fig:str}).  At $u =
\nicefrac{1}{2}$ the space group changes from $P6_{3}mc$ to
$P6_{3}/mmc$.  In the second step (\transition{HX}{RS}), the lattice
is compressed uniaxially along the $[10\bar{1}0]$ direction and
simultaneously $\nicefrac{c}{a}$ is decreased further.  The \phase{WZ} phase is
characterized by $a = b$, $\nicefrac{c}{a} = \sqrt{\nicefrac{8}{3}}$, 
$v = \nicefrac{1}{3}$ and $u = \nicefrac{3}{8}$, whereas the \phase{HX}
phase has $a = b$,  $\nicefrac{c}{a} = 1.2$, $v = \nicefrac{1}{3}$, and 
$u = \nicefrac{1}{2}$. For the
\phase{RS} phase, $a = b = c$ and $v = u = \nicefrac{1}{2}$. For
details, see Ref.~\onlinecite{Limpijumnong01b}.

To perform total-energy minimizations (see Section~\ref{sec:comp}),
the appropriate unit cell of an orthorhombic lattice with space group
$Cmc2_{1}$ was used (Fig.~\ref{fig:str}). The latter is a common
subgroup of all three phases.\cite{Sowa01} In our calculations, the
parameter $v$, specifying the relative positions of the
subsublattices,\cite{Limpijumnong01b} and the ratio $b/a$ (defined in
Fig.~\ref{fig:str}) are fixed by the geometry of a particular phase.
The parameter $u$, which is the internal vertical displacement of the
cation atoms in the oxygen plane, and the $\nicefrac{c}{a}$ ratio were
obtained by the total energy minimization for the pure compounds (i.\,e.,
\phase{WZ}-ZnO and \phase{RS}-MgO) within LSDA. They are kept fixed at these
values for \ZMO\ alloys for all concentrations $x$.

\section{Computational details}
\label{sec:comp}
The electronic and geometric structures of \ZMO\ alloys are obtained
within density-functional theory. The local density approximation and
its self-interaction correction (see below) are implemented in a
multiple-scattering approach (the Korringa-Kohn-Rostoker method,
KKR).\cite{Zabloudil05} Disordered alloys are described within the
coherent potential approximation (CPA).\cite{Gonis92}

For closed-packed systems, for example transition metals, the crystal
potential is commonly approximated as a sum over the so-called `muffin-tin
potentials'\cite{Weinberger90} (that is, spherically symmetric
potentials centered at each lattice site, while in the interstitial region
the potential is constant).  For open systems, this approximation
results in a relatively poor description of the electronic structure.
It has turned out that the usual trick of inserting the so-called `empty
spheres' into the interstitial region is not sufficient for MgO and ZnO, in
particular in view of the high degree of accuracy needed in the
evaluation of total energies.  Obviously, the latter is inevitable for
a reliable description of structural phase transitions.  Consequently,
a full-charge density approximation was applied.
%
%
Here, the total energy is estimated from the non-spherical charge
density and the full potential. All radial integrals are calculated
using the unit cell geometry; the single-site problem is still solved
with a spherical potential averaged within a Voronoi cell.  The
accuracy of this approach is as good as that of a full-potential
method but is not as time-consuming.  The validity of the full
charge-density approximation was checked by comparison with results of
the corresponding KKR full-potential calculations for the ordered
\phase{WZ}-ZnO and \phase{RS}-MgO compounds.

The disordered \ZMO\ can be viewed as a substitutional binary alloy in
which the metal sublattice (Zn, Mg) is subject to chemical disorder.
The oxygen sublattice remains unaffected. Hence, the coherent
potential approximation (CPA) is an obvious choice for describing
\ZMO\ at an arbitrary concentration $x$. Within the CPA, Mg and Zn
impurities are embedded into an effective coherent potential medium which is
determined self-consistently.\cite{Gonis92,Zabloudil05} Since the
KKR-CPA is a single-site approximation, it does not allow to
investigate the influence of chemical short-range order on the
electronic properties (
SRO can be taken into account by the cluster CPA,\cite{Weinberger88} the locally
self-consistent Green function method\cite{Abrikosov97} or the
non-local CPA approach\cite{Rowlands06}). However, for many alloys the
single-site CPA provides a reasonable description of the electronic
structure.\cite{Abrikosov98} For \ZMO, the validity of the KKR-CPA was
established through a direct comparison with KKR supercell calculations (which
include SRO effects) for $x = 0.25$, $0.50$, and $0.75$. The major
structural and electronic properties, such as the equilibrium lattice
constants, bulk moduli, equilibrium pressures, and fundamental band
gaps, are reproducible within both the KKR-CPA and the supercell
approach. Hence we conclude that SRO effects cannot be ruled out but
they are of minor importance for the issues addressed in this work.  A
detailed investigation of SRO effects is beyond the scope of the
present work.

Although the Mg-2p electron states lie comparably deeply in energy
(about \unit[2]{Rydberg} below the valence bands), they have been 
treated as valence states. It is found that the hybridization of 
the associated electronic states with the valence
states is important for an accurate determination of total energies,
especially for the evaluation of both the $\nicefrac{c}{a}$ ratio and the
parameter $u$ in the \phase{WZ} structure. The same procedure was
applied in an earlier all-electron study of MgO, using a
full-potential linearized muffin-tin orbital (FP-LMTO)
method.\cite{Limpijumnong01b}

Since the Zn-$3d$ states are energetically close to the O-$2p$ states,
they have to be considered as valence electrons as well. It is
well-known that the hybridization of the localized Zn-$3d$ electrons
with the O-$2p$ electrons is crucial for an accurate investigation of
the band-gap formation in ZnO.\cite{Usuda02} Being strongly localized,
the Zn-$3d$ states are not adequately described within the local density
approximation. The LSDA contains the (unphysical)
self-interaction of an electron with itself (see e.\,g.\
Refs.~\onlinecite{Lueders05} and \onlinecite{Temmerman98}), an effect
whose importance increases with the degree of electron localization.
As a consequence of the self-interaction, the energy levels of the
localized electrons obtained within the LSDA lie too high and their
hybridization with the O-$2p$ states is too strong.  The calculated
fundamental band gap in ZnO is therefore considerably too small as compared to
experiment.

An improved description of localized electrons is achieved by the
self-interaction correction \cite{Perdew81,Lueders05} in which
the unphysical self-interaction is removed from the LSDA exchange-correlation
functional. In this approach, SIC is applied to various
configurations of localized electron states, and the configuration
with the lowest total energy defines the ground state energy and configuration.
The SIC-LSD approach, being based on a variational principle, is parameter 
free and treats on equal footing both itinerant and localized electrons. 
When no localized electrons are present
in the system, then the SIC-LSDA total energy is equivalent to the LSDA
total energy. Thus the LSDA energy functional is a local minimum of
the SIC-LSDA functional.
For ZnO in both the \phase{RS} and the \phase{WZ} phase it is found
that all $3d$-electrons have to be SI-corrected. Since the Zn-$3d$
states have semi-core character, application of the SIC leads to a
uniform increase of the binding energies of these states, to a
decrease of the total energy, and to an increased fundamental band gap
in ZnO\@.  Because other electronic states were treated within the
LSDA, we label this calculations SIC-LSDA for short.

To check whether SIC affects only the fundamental band gap
or also structural properties, we have compared results of a SIC-LSDA 
calculation for ZnO with those of an LSDA calculation, based on the same
computer code. It has been shown that structural properties calculated
within both LSDA and SIC-LSDA are very close, whereas the band gap
obtained within SIC-LSDA has been substantially larger than in the LSDA case.
This finding has been further verified for disordered \ZMO\ alloys at
selected concentrations.
As a result, in order to save computational costs, only 
LSDA has been used for the studies of the structural properties. In
contrast, here, the fundamental band gap has been investigated at the respective
equilibrium lattice constants within SIC-LSDA\@.  For both the LSDA and
SIC-LSDA calculations the Perdew-Wang exchange-correlation
functional has been applied.\cite{Perdew92}

The equilibrium volumes, bulk moduli, pressures, and enthalpies have been
calculated for zero temperature from the total energy fitted to the
Murnaghan equation of state.\cite{Murnaghan44} Lattice vibrations,
finite-temperature effects, and relativistic corrections have not been
considered. The angular momentum cut-offs $l_{\mathrm{max}}$ have been
chosen as $3$ for the Green function expansion and $6$ for both the
charge-density and the potential representation. The convergence with
$l_{\mathrm{max}}$ and with normalization of the Green function was
significantly improved by use of Lloyd's
formula in this work.\cite{Lloyd67,Zeller04} The concentration, $x$, of Mg impurities
in ZnO has been varied in steps of $\unit[5]{\%}$.

\section{Structural properties of ordered \protect\ZNO\ and
  \protect\MGO\ in \phase{RS} and \phase{WZ} phases}
\label{sec:struct-prop-order}
Due to the large number of structural parameters, a complete structure
optimization of \ZMO\ alloys is an involved task. Apart from these
parameters, the alloy composition $x$ is an additional degree of
freedom which complicates the problem further. Therefore, the number
of parameters to be optimized has been reduced by concentrating on volume
changes in the \phase{WZ}, \phase{HX}, and \phase{RS} structures upon
variation of $x$ (see Section~\ref{sec:struct-phase-trans}).

To our knowledge, the KKR method was not used before for optimizing
\phase{WZ} and \phase{HX} structures. Hence extensive calculations
for pure ZnO and MgO have been required to determine both the optimum $u$
and $\nicefrac{c}{a}$. In accomplishing this, the procedure suggested in
Ref.~\onlinecite{Limpijumnong01b} has been followed. Since the calculation of
lattice relaxations from forces is rather complicated and not sufficiently
accurate within the KKR method, we have calculated the total energies 
consecutively varying three parameters: the lattice constants $a$, the
$\nicefrac{c}{a}$ and the internal parameter $u$. The total energy has
been calculated at a given $u$ for a sufficiently dense mesh of lattice 
constants and then fitted to the Murnaghan equation of state. The initial value
of the parameter $u$ has been taken from the experiment for a particular
structure and then successively varied as long as the absolute total
energy minimum has not been reached.  To establish the validity of the 
present KKR approach for structure optimization, our results have been
compared with those of other first-principles calculations.  We note
that our calculations for both ZnO and MgO in various
structures have been carried out for the same unit cell on the same level
of approximation.

The adequacy of the present approach is evidenced by the agreement with
experimental data and the results of other theoretical approaches
(see Tables~\ref{tab:ZnO} and \ref{tab:MgO}). It is found that our results
agree with those obtained by the other all-electron methods, as opposed
to the pseudo-potential methods. In particular, the parameters obtained in
this work compare well with those reported by Limpijumnong and
coworkers.\cite{Limpijumnong01b,Limpijumnong04} This agreement might
be attributed to the fact that both studies follow the same
optimization scheme (as suggested in
Refs.~\onlinecite{Limpijumnong01b} and \onlinecite{Limpijumnong04})
and that the Zn-$3d$ as well as the Mg-$2p$ electrons have been treated as
valence electrons.

\begin{table*}
  \centering
  \caption{Properties of ZnO in the wurtzite (\phase{WZ}), the hexagonal
    (\phase{HX}), and the rock-salt (\phase{RS}) phase.
    Equilibrium volumes $V_{0}$, $\nicefrac{c}{a}$ ratios, internal
    parameters $u$, and bulk moduli $B_{0}$ are compiled for theoretical
    (present work displayed in bold) and for experimental works.
    PPW, LCAO, and FP-LMTO are short for 
    pseudo-potential plane waves, linear combination of atomic orbitals, and
    full-potential linearized muffin-tin orbital, respectively.}
  \label{tab:ZnO}
  \begin{ruledtabular}
    \begin{tabular}{llll}
      Phase
      &
      & \multicolumn{1}{c}{Theory} 
      & \multicolumn{1}{c}{Experiment}
      \\
      \hline
      \phase{WZ}
      &
      $V_{0}$ (\AA$^3$)
      & 
      22.80\footnote[1]{Ref.~\onlinecite{Limpijumnong04}: PPW.},
      \textbf{22.83},
      22.87\footnote[2]{Ref.~\onlinecite{Jaffe00}: LCAO.},
      22.91\footnote[3]{Ref.~\onlinecite{Uddin06}: LCAO.},
      22.93\footnote[4]{Ref.~\onlinecite{Seko05}: PPW.},
      23.4\footnote[5]{Ref.~\onlinecite{Kim01b}:
        PPW.}\footnote[6]{Ref.~\onlinecite{Lambrecht00}: FP-LMTO.},
      23.62\footnote[7]{Ref.~\onlinecite{Ahuja98}: FP-LMTO.},
      23.78\footnotemark[5]
      &
      23.80  \footnote[8]{Ref.~\onlinecite{Desgreniers98}.}\footnote[9]{Ref.~\onlinecite{Karzel96}.}
      \\
      &
      $\nicefrac{c}{a}$  
      &  1.590\footnotemark[7],
      \textbf{1.602},
      1.605\footnotemark[5],
      1.607\footnotemark[3],
      1.608\footnotemark[6],
      1.610\footnotemark[1],
      1.614\footnotemark[2],
      1.617\footnotemark[4]
      &
      1.602\footnotemark[8]\footnotemark[9]
      \\ 
      &
      $u$  
      & 0.379\footnotemark[2],
      0.380\footnotemark[1]\footnotemark[7]\footnotemark[6],
      \textbf{0.381}\footnotemark[3]
      & 
      0.382\footnotemark[8]\footnotemark[9]
      \\
      & 
      $B_{0}$(\unit{GPa})  
      &
      \textbf{154}, 154\footnotemark[5],
      155\footnotemark[3],
      157\footnotemark[6],
      160\footnotemark[7],
      162\footnotemark[1]\footnotemark[2]\footnotemark[4]
      &
      143\footnotemark[8],
      183\footnotemark[9] \\ 
      \hline
      \phase{HX}
      &
      $V_{0}$ (\AA$^3$)  &  
      \textbf{22.12} \\
      &
      $\nicefrac{c}{a}$ 
      & \textbf{1.200} 
      \\
      &
      $B_{0}$(\unit{GPa}) 
      & \textbf{165} \\
      \hline
      \phase{RS}
      &
      $V_{0}$ (\unit{\AA$^{3}$})  &  
      18.70\footnotemark[1],
      18.76\footnotemark[3],
      18.87\footnotemark[4],
      \textbf{18.88},   
      18.98\footnotemark[2],
      19.08\footnotemark[7],
      19.45\footnotemark[5]
      &
      19.48\footnotemark[9],
      19.60\footnotemark[8]
      \\
      &
      $B_{0}$(\unit{GPa})  &
      200\footnotemark[5],
      \textbf{201},
      203\footnotemark[3],
      206\footnotemark[2],
      210\footnotemark[1],
      211\footnotemark[4],
      219\footnotemark[7]
      &
      202\footnotemark[9],
      228\footnotemark[8]
    \end{tabular}
  \end{ruledtabular}
\end{table*}

\begin{table*}
  \centering
  \caption{As Table~\ref{tab:ZnO}, but for MgO\@. Missing references
    are given in Table~\ref{tab:ZnO}.}
  \label{tab:MgO}
  \begin{ruledtabular}
    \begin{tabular}{llll}
      Phase
      &
      & \multicolumn{1}{c}{Theory} &  \multicolumn{1}{c}{Experiment}  \\
      \hline
      \phase{WZ}
      &
      $V_{0}$ (\unit{\AA$^{3}$})
      &  
      22.50\footnote[10]{Ref.~\onlinecite{Limpijumnong01b}: FP-LMTO.},,
      22.53\footnotemark[5],
      23.15\footnotemark[6],
      \textbf{23.20}
      & \\
      &
      $\nicefrac{c}{a}$ &
      1.550\footnotemark[5], 
      \textbf{1.610},
      1.620\footnotemark[10],
      1.633\footnotemark[6] & \\
      &
      $u$ &
      \textbf{0.380}, 0.380 \footnotemark[10] & \\
      &
      $B_{0}$(\unit{GPa}) &
      \textbf{121},
      131\footnotemark[5], 137\footnotemark[10] & \\
      \hline
      \phase{HX}
      &
      $V_{0}$ (\unit{\AA$^{3}$})  
      &  
      20.90\footnotemark[10], \textbf{21.71} 
      &
      \\
      &
      $\nicefrac{c}{a}$ 
      &  
      \textbf{1.200}, 1.200\footnotemark[10] 
      &
      \\
      &
      $B_{0}$(\unit{GPa})  
      & 
      \textbf{135}, 148\footnotemark[10] 
      &
      \\ 
      \hline
      \phase{RS}
      &
      $V_{0}$ (\unit{\AA$^{3}$})  &  
      17.54\footnotemark[5],
      17.80\footnotemark[10],
      17.96\footnotemark[4],
      18.03\footnotemark[2],
      \textbf{18.19},
      18.65\footnote[11]{Ref.~\onlinecite{Baranov02}: FP-KKR.} &
      18.67\footnote[12]{Ref.~\onlinecite{Wyckoff63}.},
      18.75\footnote[13]{Ref.~\onlinecite{Sangster81}.} \\ 
      &
      $B_{0}$(\unit{GPa})  &
      \textbf{167},
      170\footnotemark[5],
      172\footnotemark[11],
      174\footnotemark[4],
      178\footnotemark[10],
      186\footnotemark[2] 
      &
      169\footnote[14]{Ref.~\onlinecite{Cohen76}.},
      172\footnote[15]{Ref.~\onlinecite{Sangster70}.}
    \end{tabular}
  \end{ruledtabular}
\end{table*}

As for the structural properties, we find agreement concerning the 
volume ratios of the
different phases and the equilibrium pressures at the phase
transitions (see Tables~\ref{tab:phaseZnO} and~\ref{tab:phaseMgO}).
For ZnO there is only the \transition{WZ}{RS} phase transition while
for MgO there are two transitions, namely \transition{RS}{HX} and
\transition{RS}{WZ}.  In particular for ZnO, our theoretical results
compare well with experimental ones (see Table~\ref{tab:phaseZnO}).
\begin{table}
  \centering
  \caption{Structural phase transition in ZnO\@. Given are the volume ratio and 
    the equilibrium pressure at the \transition{WZ}{RS} phase transition. 
    References as in Table~\ref{tab:ZnO}. Present work displayed in bold.}
  \label{tab:phaseZnO}
  \begin{ruledtabular}
    \begin{tabular}{lll}
      & \multicolumn{1}{c}{Theory} &  \multicolumn{1}{c}{Experiment}  \\
      \hline
      $V_{\mathrm{\phase{WZ}}} / V_{\mathrm{\phase{RS}}}$
      & 1.20\footnotemark[2], \textbf{1.21},
      1.22\footnotemark[3]\footnotemark[4]\footnotemark[5], 1.24\footnotemark[7]
      & 1.21\footnotemark[9], 1.22\footnotemark[8]\\
      $P_{\transition{WZ}{RS}}$ (\unit{GPa})         
      & 3.9\footnotemark[3], 6.6\footnotemark[3], 8.0\footnotemark[7],
      8.2\footnotemark[1], \textbf{8.6}, 8.7\footnotemark[3]  &
      8.7\footnotemark[8], 9.1\footnotemark[9]  
    \end{tabular}
  \end{ruledtabular}
\end{table}
For MgO we find agreement with other theoretical estimations too
(Table~\ref{tab:phaseMgO}). However, the equilibrium pressure for the
\phase{RS}-\phase{HX} phase transition might be an exception. The
apparent disagreement with the result reported in
Ref.~\onlinecite{Limpijumnong01b} may be related to the instabilities
of the \phase{WZ} and the \phase{HX} structures
(Figs.~\ref{fig:enthalpy} and~\ref{fig:energy}a).  Moreover, due to 
the negative pressure these phases cannot be realized experimentally,
hence ruling out a clarifying comparison of theory with experiment.
In favor of our work, we would like to mention that the present
calculations reproduce well $\nicefrac{c}{a}$ and $u$. One might
speculate that earlier implementations of multiple-scattering theory
failed to optimize these structural parameters in any open structure
due to the slow angular momentum convergence of the Green function
(see Section~\ref{sec:comp}; the convergence is significantly improved
by using Lloyd's formula).\cite{Moghadam01}
\begin{table}
  \centering
  \caption{As Table~\ref{tab:phaseZnO}, but for MgO\@.
    For MgO there are two transitions, \transition{RS}{HX} and \transition{RS}{WZ}.
    Experimental data are not available. References as in Table~\ref{tab:MgO}. 
    Present work displayed in bold.}
  \label{tab:phaseMgO}
  \begin{ruledtabular}
    \begin{tabular}{ll}
      & \multicolumn{1}{c}{Theory} \\
      \hline
      $V_{\mathrm{\phase{HX}}} / V_{\mathrm{\phase{RS}}}$
      & 1.17\footnotemark[10], \textbf{1.19} 
      \\
      $P_{\transition{RS}{HX}}$ (\unit{GPa})  
      &
      $-16.2$\footnotemark[10], $\mathbf{-8.5}$
      \\
      \hline
      $V_{\mathrm{\phase{WZ}}} / V_{\mathrm{\phase{RS}}}$
      & \textbf{1.24}, 1.26\footnotemark[10], 1.28\footnotemark[5]
      \\
      $P_{\transition{RS}{WZ}}$ (\unit{GPa})         
      & $\mathbf{-11.1}$, $-8.4$\footnotemark[10]
    \end{tabular}
  \end{ruledtabular}
\end{table}

In summary, the agreement of our results for the ordered
alloys with those of other theoretical investigations and experiments
indicates that the present approach is as reliable and
accurate as any other state-of-the-art methods.

\section{Structural properties of disordered \protect\ZMO\ alloys}
\label{sec:struct-prop-disorder}
Having established that our results for ordered ZnO and MgO are
consistent with the structural properties obtained by other
theoretical methods and by experiments, we now turn to the discussion
of results for disordered \ZMO\ alloys.  To reduce the number of
parameters to be optimized, both $\nicefrac{c}{a}$ and $u$ were fixed
respectively to $1.6$ and $0.38$ in the \phase{WZ} structure and to $1.2$ 
and $0.5$ in the \phase{HX} structure. This approximation
is justified by the weak dependence of both $\nicefrac{c}{a}$ and $u$
on the atomic species (Mg or Zn; cf.\ Tables~\ref{tab:ZnO}
and~\ref{tab:MgO}).  Consequently, the structure (the actual phase), 
Mg concentration $x$, and unit cell volume remain to be varied. The results
of the total-energy calculations involving these three variables 
are discussed below. The results of the total energy
calculations are comprised in the formation enthalpy (at T=0)
\begin{align}
  \label{eq:1}
  \begin{aligned}
    \Delta H_{\alpha}(\mathrm{Mg}_{x}\mathrm{Zn}_{1-x}\mathrm{O})
    & =
    E_{\alpha}(\mathrm{Mg}_{x}\mathrm{Zn}_{1-x}\mathrm{O})
    \\
    - x E_{\mathrm{\phase{RS}}}(\mathrm{MgO})
    &
    - (1-x) E_{\mathrm{\phase{WZ}}}(\mathrm{ZnO}) 
  \end{aligned}
\end{align}
of a structure $\alpha$ of \ZMO, relative to the most stable forms of
ZnO (\phase{WZ}) and MgO (\phase{RS}) compounds.\cite{Sanati03b}

\begin{figure}
  \centering
  \includegraphics[width=0.45\textwidth]{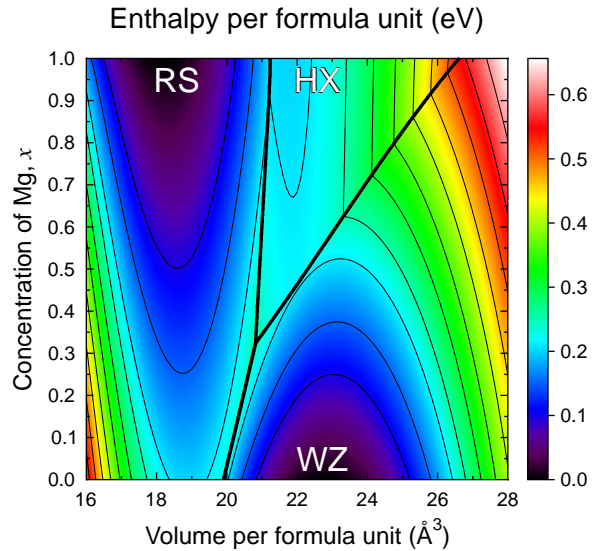}
  \caption{(Color) Formation enthalpy of \ZMO\ alloys versus
    equilibrium volume and Zn concentration, depicted as color scale
    [cf.\ eq.~(\ref{eq:1})].  Straight bold lines mark separations
    between the phases (\phase{WZ}, \phase{HX}, and \phase{RS}).}
  \label{fig:enthalpy}
\end{figure}

The formation enthalpy is positive throughout, as seen in
Fig.~\ref{fig:enthalpy}, in agreement with previous theoretical
studies.\cite{Sanati03b,Seko05} There are two global minima (i.\,e.\
the most stable phases; black, dark regions), for pure ZnO in
\phase{WZ} ($x = 0$) and pure MgO in \phase{RS} phase ($x = 1$). This
finding implies a tendency towards phase separation if the integration
of different constituents into the medium cannot be
maintained.\cite{Sanati03b}

In the following, five special cases are discussed in more detail, namely
$x = 0.0$, $0.33$, $0.52$, $0.71$, and $1.0$. 

\begin{figure}
  \centering
  \includegraphics[width=0.45\textwidth]{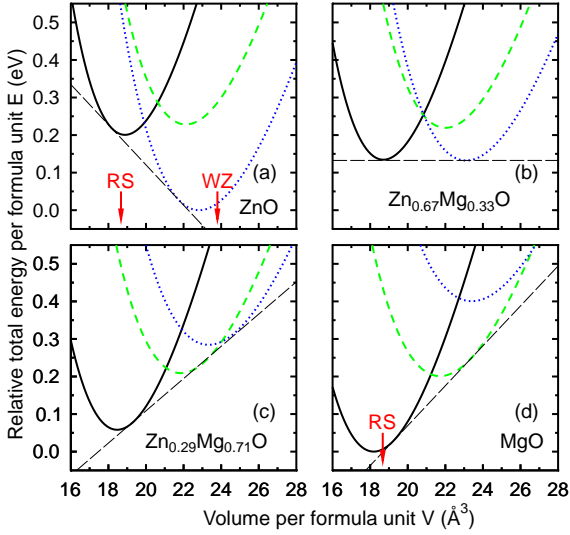}
  \caption{(Color online) Relative total energies of \ZMO\ alloys in
    \phase{RS} (black solid line), \phase{HX} (green dashed line), and
    \phase{WZ} (blue dotted line) structures at four different
    concentrations: (a) $x = 0.0$ (pure ZnO), (b) $x = 0.33$, (c) $x =
    0.71$, and (d) $x = 1.0$ (pure MgO). Total energies in cases (a)
    are taken relative to the \phase{WZ} phase at $x = 0.0$ and, in
    (b), (c), and (d), relative to the \phase{Mg} phase at $x = 1.0$. 
    Red arrows show experimental values of volume (see
    Tables~\ref{tab:ZnO} and \ref{tab:MgO})}
  \label{fig:energy}
\end{figure}

\begin{figure}
  \centering
  \includegraphics[width=0.45\textwidth]{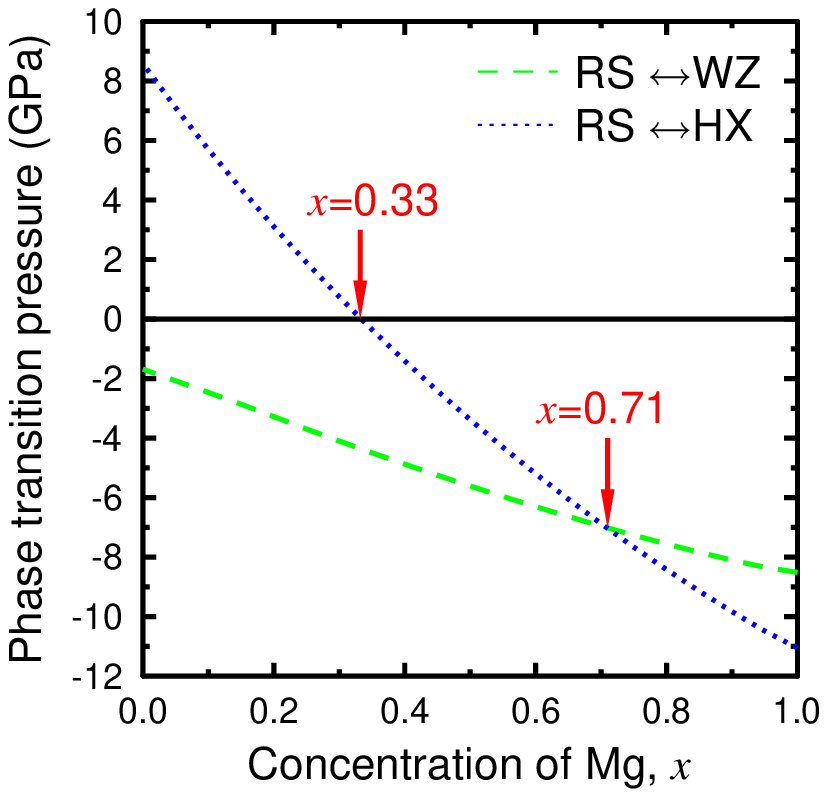}
  \caption{(Color online) Equilibrium pressure in \ZMO\ alloys:
    \transition{RS}{WZ} (green dashed line) and \transition{RS}{HX}
    (blue dotted line).}
  \label{fig:pressure}
\end{figure}

With low Mg concentration, the \phase{WZ} phase is favorable in
any case (Fig.~\ref{fig:energy}a). The pressure needed for a
\transition{WZ}{RS} transition is positive and increases monotonously
(Fig.~\ref{fig:pressure}).


\begin{figure}
  \centering
  \includegraphics[width=0.45\textwidth]{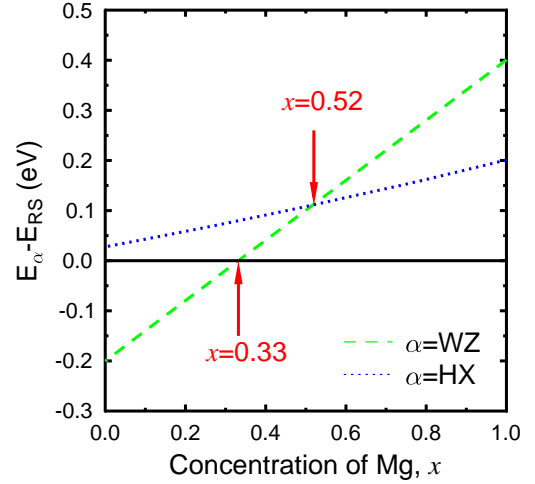}
  \caption{(Color online) Total energies of \ZMO\ alloys in the
    \phase{HX} (blue dashed line) and the \phase{WZ} (green dotted
    line) structures, taken relative to the total energy of the
    \phase{RS} structure at equilibrium volumes.}
  \label{fig:de}
\end{figure}

For $x < 0.33$ the total energy of the \phase{HX} structure is larger
than for the \phase{RS} and the \phase{WZ} phases, implying that an
intermediate \phase{HX} phase cannot be established
(Fig.~\ref{fig:enthalpy}). Consequently, a direct \transition{WZ}{RS}
transition is possible at positive pressure. At $x = 0.33$
(Figs.~\ref{fig:energy}b and~\ref{fig:de}) this phase transition can
take place at zero pressure.  This finding is consistent with the
theoretical work of Sanati and coworkers\cite{Sanati03b} and is also
observed experimentally in ZnO-MgO heterostructures.\cite{Ohtomo98}

At $x = 0.71$ the \transition{WZ}{HX} and the \transition{HX}{RS}
transitions occur at the same pressure (Fig.~\ref{fig:pressure}). This
is evident from Fig.~\ref{fig:energy}c because all total energy curves
can be connected by a single tangent.  We note that for $x = 0.52$ the
\transition{WZ}{HX} transition is possible at zero 
pressure, and the corresponding total energies relative to \phase{RS}
phase are identical (see Fig.~\ref{fig:de}).

For pure MgO ($x = 1.0$;
Fig.~\ref{fig:energy}d), the \phase{RS} structure exhibits the global
minimum at a volume of $\unit[18.19]{\AA^{3}}$ which is about
\unit[2]{\%} less than the experimental value (see
Table~\ref{tab:MgO}).  By applying a negative pressure (i.\,e.\ by
increasing the volume along the tangent) the \phase{RS} phase is
transformed into the \phase{HX} structure, in agreement with the work
of Limpijumnong and coworkers.\cite{Limpijumnong01b}

According to the phase diagram (Fig.~\ref{fig:enthalpy}), the
\phase{HX} phase is an intermediate phase between the \phase{WZ} and
\phase{RS} structure for $0.33 \le x \le 1.00$.  Although the
\transition{HX}{RS} transition pressure increases with Mg content, it
remains negative in the whole range of concentrations
(Fig.~\ref{fig:pressure}).

\section{Band gap in \protect\ZMO\ alloys}
\label{sec:band-gap-engineering}
Since \ZMO\ alloys arouse great interest as band gap-engineering
materials, in this section we study the evolution of their fundamental 
band gap as a function of the Zn and Mg contents. We restrict our consideration
to the \phase{RS} and \phase{WZ} phases, as only these appear
interesting from the experimental point of view.

Band gap formation can be correctly described only within a many-body
theory which takes properly into account the electron-hole excitations.
However, the $GW$ approximation\cite{Hedin69}, probably the most
popular first-principles many-body approach for calculating excitation energies, 
is too expensive for studying alloys with arbitrary concentrations and
also having constituents with localized electrons. Therefore, 
for this study we have used only the LSDA and SIC-LSDA approaches. 
The calculated band gaps have been estimated from the density of states 
(DOS), band structure and Bloch spectral function, corresponding to
the theoretical equilibrium lattice structure of a given chemical composition.

\begin{figure}
  \centering
  \includegraphics[width=0.95\columnwidth]{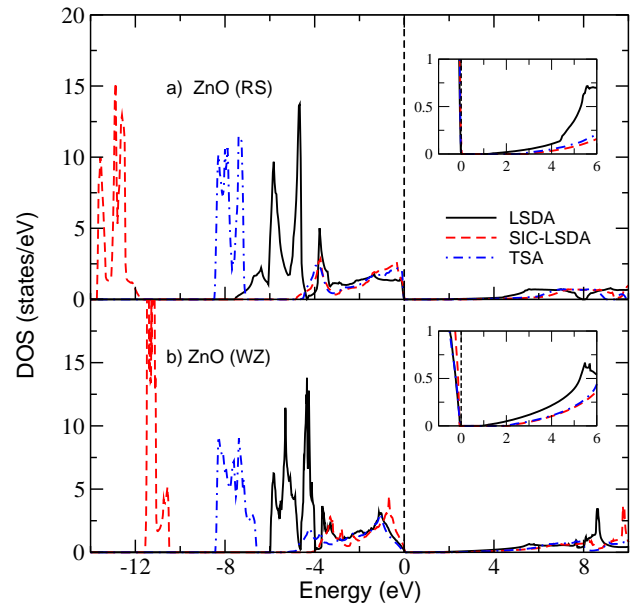}
  \caption{(Color) Density of states (DOS) of ZnO in the
    \phase{RS} (a) and the \phase{WZ} (b) phase obtained with the LSDA
    (black solid lines), the SIC-LSDA (red dashed lines) and SIC-TSA
    (blue dash-dotted lines) approaches.
    Insets show the behaviour of the DOS and fundamental band gap
    close to the Fermi level. The energy is given relative the Fermi
    level.}
  \label{fig:dos}
\end{figure}

Using LSDA for \phase{RS}-MgO we obtain a band gap of $\unit[5.15]{eV}$,
which is $\unit[67]{\%}$ of the experimental value ($\unit[7.7]{eV}$).
For the pure \phase{WZ}-ZnO, LSDA gives the band gap of $\unit[0.8]{eV}$,
which is only $\unit[23]{\%}$ of the experimental value of $\unit[3.35]{eV}$.
In MgO, the difference between the calculated and experimental values
originates mostly from the fact that the electron-hole excitations are 
neglected in LSDA, which can be corrected effectively with the $GW$
approximation.\cite{PhysRevB.52.8788} For ZnO the problem is more complex,
as in addition to the neglect of the electron-hole excitations, the localized
nature of Zn 3$d$ electrons is not adequately represented within LSDA. This
failure of LSDA is largerly related to the inherent unphysical self-interaction.
The latter affects the band gap by placing the localized 3$d$ electrons
of Zn at too low binding energies, thus leading to their strong
hybridization with the O-$2p$ states. In LSDA the Zn-$3d$ states
are about $\unit[3]{eV}$ too high with respect to the experimental
value of about $\unit[7.5]{eV}$--$\unit[8.5]{eV}$\cite{PhysRevLett.27.97,PhysRevB.5.2296}
(see the DOS of ZnO in Fig.~\ref{fig:dos}).
In the \phase{RS} and \phase{WZ} phases, the $3d$ states are placed at slightly different
energies, and in the \phase{RS} the corresponding band width is larger
due to the more closed-packed crystal environment. The strong
$pd$-hybridization dramatically reduces the band gap, which can be
improved only slightly within the $GW$ approximation, if based
on the LSDA Green's function.\cite{Usuda02} It is the aim of our future studies to 
use the SIC-LSDA band structure of ZnO for the subsequent $GW$ calculation. 
The SIC-LSDA band gap we have calculated is $\unit[2.42]{eV}$ for \phase{WZ}-ZnO
and $\unit[2.96]{eV}$ for \phase{RS}-ZnO. Thus, the SIC-LSDA band gap for
the \phase{WZ}-ZnO constitutes 69\% of the experimental value of $\unit[3.35]{eV}$.
This tells us that SIC-LSDA is at least as good for the pure
ZnO, as LSDA is for the pure MgO. The reason being, that SIC-LSDA describes
both itinerant and localized electrons on equal footing. Of course, in MgO 
the band gap is constituted by the $sp$ electrons which are itinerant and thus 
unaffected by the spurious self-interaction, while in ZnO the 3$d$ electrons of 
Zn play a defining role in establishing the band gap. The SIC-LSDA approach,
by removing the unphysical self-interaction of all 10 Zn $d$ electrons, describes 
them much more adequately than LSDA.
%
\begin{figure}
  \centering
  \includegraphics[width=0.9\columnwidth]{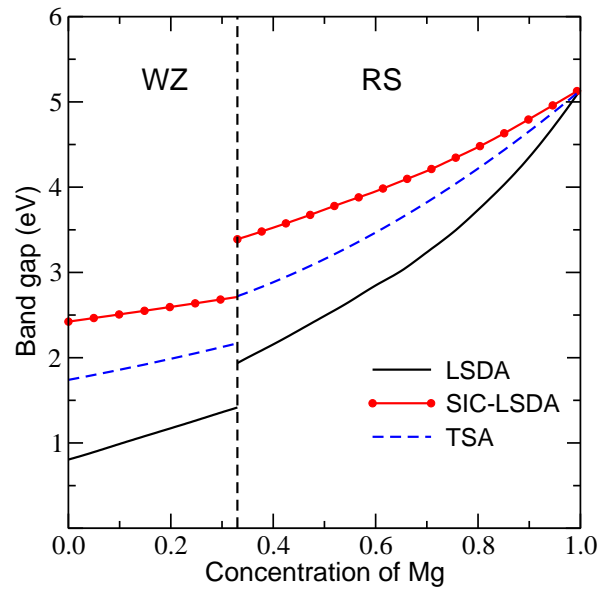}
  \caption{(Color online) Band gaps of \ZMO\ alloys in the \phase{RS}
    and \phase{WZ} phases versus Mg content, calculated within LSDA and
    SIC-LSDA (line styles as indicated).}
  \label{fig:gap}
\end{figure}
What happens as a result of SIC is that all the $d$-states of Zn move to
higher binding energies and the Zn-derived bands become narrower thereby reducing the
$pd$-hybridization and increasing the band gap, as seen in Fig.~\ref{fig:dos}.
The structural properties of \ZMO\ are believed to be little affected by 
this uniform shift downwards in energy of the $d$-states, leading mostly to 
a uniform lowering of the total energy. 

Being an effective one-electron ground state theory, 
SIC-LSDA does not provide a quasi-particle spectrum to compare with
spectroscopies. Missing the crucial screening/relaxation effects (self-energy),
it predicts the 3$d$ Zn electrons at too high binding energies, as opposed
to LSDA where they come out too low (Fig.~\ref{fig:dos}). One can implement
a simple fix to correct for the screening effects in SIC-LSDA, based on 
the Slater's transition state theory.\cite{Slater72,Liberman00} Following the concept of 
the latter, we calculate the SIC-LSDA-based removal energies of localized electrons 
as
%
%
%
an average of the calculated SIC-LSDA and LSDA $d$-state expectation values
\begin{align}
 \label{eq:2}
\varepsilon_{TSA}=\frac{1}{2}(<d|H_{LSDA}+V_{SIC}|d>+<d|H_{LSDA}|d>)\,.
\end{align}
Effectively, the above 
equation states that only half of the SIC potential should be applied 
at the stage of calculating the density of states, after the self-consistency
has been achieved. We refer to Eq.~\ref{eq:2} as the transition state approximation
(TSA) and show the resulting DOS of ZnO in Fig.~\ref{fig:dos}.
%
We can see that Zn $d$-states, calculated using TSA, appear at lower binding
energies, as compared to the strict SIC-LSDA result. Consequently, the TSA-binding 
energies of Zn $d$ states are in better agreement with experimental values of
$\unit[7.5]{eV}$--$\unit[8.5]{eV}$\cite{PhysRevLett.27.97,PhysRevB.5.2296}.
The effective hybridization of Zn $d$-states with the oxygen $p$-states
is stronger as in the SIC-LSDA case, which leads to a reduction of the
band gap to $\unit[2.32]{eV}$ and $\unit[1.74]{eV}$~(52\% of the
experimental value) in \phase{RS} and
\phase{WZ} phases, respectively. 

In \ZMO\ alloys the size of the band gap depends on the
concentration of Mg impurities, as seen in Fig.~\ref{fig:gap} and Table \ref{tab:gap}. 
While at the Mg rich end the LSDA band gaps are closer to experiment, 
at the other end the SIC-LSDA band gaps are more adequate. As for the TSA results,
they fall mostly in between the LSDA and SIC-LSDA band gaps, especially for 
small Mg concentrations, where the hybridization of the O 2$p$ bands with the Zn 3$d$ states
is of great significance (see Fig.~\ref{fig:gap}). The behaviour of the band gap as a function 
of Mg concentration is also different between the various approaches. While the LSDA curves are 
rather parabolic, the SIC-LSDA and TSA band gaps seem to change almost linearly with
concentration, thus following more closely experimental results\cite{Chen04,Schmidt-Grund05}. 
This is mostly due to the fact that the actual magnitudes of the band gaps change 
slower with concentration in SIC-LSDA and TSA than in LSDA. The reason being that the
larger Zn content, the more inadequate LSDA is and the smaller the resulting band gaps.   
%
\begin{table}
  \centering
  \caption{The band gap $E{_g}$ for MgO (\phase{RS}), ZnO(\phase{WZ}) and
    the bowing parameter for \ZMO\ alloys.}
  \label{tab:gap}
  \begin{ruledtabular}
    \begin{tabular}{lcccc}
      & LSDA & TSA & SIC-LSDA & Experiment  \\
      \hline
      &      
\multicolumn{4}{c}{Band gap [eV]}   \\
      \hline
      \phase{RS}-MgO & 5.15 &      &      & 7.70  \\
      \phase{WZ}-ZnO & 0.80 & 1.74 & 2.42 & 3.35 \\
      \hline
      &   \multicolumn{4}{c}{ Bowing parameter [eV]}    \\
      \hline
         \phase{RS}    & 1.90 & 0.91 & 0.59 & 0.70$\pm$ 0.2 \cite{Chen04}\\
         \phase{WZ}    & 4.88 & 4.05 & 3.65 & 3.60$\pm$ 0.6 \cite{Schmidt-Grund05}
    \end{tabular}
  \end{ruledtabular}
\end{table}

Although, both LSDA and SIC-LSDA underestimate the size of the band gap, $E_g$,
its dependence on the Mg concentration, $x$, can
be compared with experimental results. One way of doing it
is to estimate the so-called bowing parameter, $b$,
appearing in the commonly used definition of the fundamental band gap
dependence on the composition $x$, namely
\begin{equation}
\label{eq:bow:1}
E_{\mathrm{g}} (x) =  x E_{\mathrm{g}} ({\mathrm{MgO}}) 
+ (1 - x) E_{\mathrm{g}} ({\mathrm{ZnO}})
- b x (1 - x), \nonumber
\end{equation}
where the bowing parameter is given by
\begin{equation}
\label{eq:bow:2}
b=2 E_{\mathrm{g}} ({\mathrm{MgO}}) + 2 E_{\mathrm{g}}
({\mathrm{ZnO}}) -4 E_{\mathrm{g}} ({\mathrm{Zn}}_{0.50}
\mathrm{Mg}_{0.50}\mathrm{O}).
\end{equation}
Thus to evaluate the bowing parameter one needs to know the band gaps 
of the pure MgO(\phase{RS}) and ZnO(\phase{WZ}), as well as their alloy 
with the concentration $x=0.5$. 

According to our total energy calculations the \ZMO\ at $x=0.5$ 
occurs in the \phase{RS} phase which is in agreement with the experiment
of Chen {\em et al.}.\cite{Chen04} However, the crystal
structure of \ZMO\ thin film alloys depends strongly on the growth
method. In variance to Ref.~\onlinecite{Chen04}, the \ZMO\ films
with $x\le 0.53$, prepared with the PLD procedure, were found to have
the \phase{WZ} structure. Therefore, for direct comparison with
the experiments  we have used Mg$_{0.50}$Zn$_{0.50}$O in both \phase{RS}
and \phase{WZ} phases. The resulting calculated bowing parameters for  
\ZMO\ alloys are presented in Table \ref{tab:gap}. We can see that the 
LSDA systematically overestimates the bowing parameter for both structures,
while the straight SIC-LSDA agrees very well with experiments. The TSA
results fall in between the LSDA and SIC-LSDA values.

Summarizing this section, we have to say that despite rather good agreement
of our SIC-LSD bowing parameter with experiments, to be truly predictive,
one would need a more robust method like a combination of SIC and $GW$. This
way we could also predict the correct magnitudes of the band gaps and enter 
the serious business of band gap engineering.


\section{Conclusions and outlook}
\label{sec:conclusions-outlook}
Structural phase transitions and the fundamental band gaps of \ZMO\
alloys have been investigated by detailed first-principles
calculations. The multiple-scattering theoretical approach used here
(Korringa-Kohn-Rostoker method) allows to treat disordered alloys
within the coherent potential approximation, that is \ZMO\ alloys with
arbitrary Zn concentration $x$, with an accuracy as good as in other
first-principles methods. The importance of treating
localized states of ZnO appropriately within the framework of the
local-density approximation to density-functional theory is
established, thereby confirming the usefulness of
the self-interaction correction.

The delicate interplay of geometry and electronic structure is not
only of importance for bulk systems, as shown in this work. In
nanotechnology, interfaces and surfaces play an essential role.
Therefore, a correct description of disordered alloys and their
electronic structure, in particular at the Fermi energy, is necessary
for predicting material properties, besides confirming and explaining
experimental results. However, to be fully predictive regarding the 
properties of such systems as \ZMO\ alloys, of importance for device applications, 
one needs a first-principles approach like a combination of $GW$ with
SIC-LSDA.

\acknowledgments

This work is supported by the \textit{Sonderforschungsbereich} SFB 762,
"Functionality of Oxidic Interfaces". Research at the Oak Ridge
National Laboratory was sponsored by the Division of Materials
Sciences and Engineering, Office of Basic Energy Sciences, US
Department of Energy, under Contract DE-AC05-00OR22725 with
UT-Battelle, LLC.  We gratefully acknowledge H. L. Meyerheim for many
stimulating discussions. The calculations were performed at the John
von Neumann Institute in J\"ulich and Rechenzentrum Garching of the
Max Planck Society (Germany). 
%
%
\bibliographystyle{apsrev}
\bibliography{./ZnMgO}
\end{document}